# Battery Capacity of Deferrable Energy Demand

Daria Madjidian, Mardavij Roozbehani, and Munther A. Dahleh

*Abstract*—We investigate the ability of a homogeneous collection of deferrable energy loads to behave as a battery; that is, to absorb and release energy in a controllable fashion up to fixed and predetermined limits on volume, charge rate and discharge rate. We derive bounds on the battery capacity that can be realized and show that there are fundamental trade-offs between battery parameters. By characterizing the state trajectories under scheduling policies that emulate two illustrative batteries, we show that the trade-offs occur because the states that allow the loads to absorb and release energy at high aggregate rates are conflicting.

## I. INTRODUCTION

The power systems is unique in that generation must meet demand at all times and in the face of uncertainty. To meet this requirement, the system operator relies on balancing reserves, which are often provided by gas power plants or fossil fuel based spinning generators. However, following the recent and rapid developments in communication and metering technology, flexible electricity loads have become viable candidates for providing low-cost and fast-acting balancing services to the power system [1], [2].

Many methods have been suggested for providing demand-based balancing support. An early example can be found in [3], where dispersed loads are equipped with governor-like controllers that respond to changes in system frequency. More recent approaches focus on controlling the aggregate consumption of a collection of loads. This includes work on thermostatically controlled loads [4]–[6], HVAC systems [7], electrical vehicle charging [8], [9], and residential pool pumps [10]. For a comprehensive survey of demand-side balancing support we refer to [1].

Despite the large interest in demand-side balancing services, few attempts have been made to quantify their *capacity*, that is, the set of power adjustments that can be tolerated without causing unacceptable disruption to any of the end-users. The focus of most methods in the literature is not on designing power scheduling policies that are robust to a specific set of power adjustments, but rather, on tracking an arbitrary trajectory. A drawback is that additional effort is required to determine the resulting capacity; for instance, by using the method in [11]. A second and more severe limitation is that the associated capacity under such tracking policies is *fixed* and cannot be adapted to changes in power system operating conditions.

In this paper, we investigate the ability of a collection of deferrable energy loads (i.e., energy demand that can be postponed up to a deadline) to directly *emulate a battery*, that is, to track any fluctuations in aggregate power supply that are compatible with the charge/discharge behavior of the battery. We define the capacity of a battery in terms of three parameters: the volume of energy that it can store, its maximum charge rate, and its maximum discharge rate. Our goal is to quantify the battery capacity that can be offered. To provide an insightful answer, we study this question for a collection of homogeneous loads with periodic arrivals.

Our contributions are as follows. We provide upper bounds on the capacity of the batteries that can be emulated and show that there are substantial trade-offs between battery parameters. We then show that these trade-offs occur because the initial energy allocation required to track large positive deviations in aggregate power supply is different than the energy allocation required to follow large negative deviations. Since the dynamics of the loads prevent instantaneous transitions between different energy allocations, and since the aggregate power adjustment is not known beforehand, there is a fundamental trade-off between the abilities of the collective load to absorb and release energy at high aggregate rates. This trade-off is not an artifact of our homogeneous load model, and is expected to occur in far more general settings.

*Related work*

Battery models have previously been used in the literature, for example, in [6], [11], [12], to quantify the capacity of a collection of flexible loads. Our work is conceptually closely related to [6], where the authors derive both upper and lower bounds on the battery capacity that can be offered by a collection of thermostatically controlled loads. An important difference is that while a deferrable load is only able to provide temporary energy storage, each thermostatically controlled load can act as a battery on its own. The problem in [6] is to aggregate a collection of heterogeneous batteries into a larger battery, and in our work it is to aggregate temporary and overlapping storage capacities into a

This work was supported by the NSF CPS Collaborative Project Award 1135843, the MIT–Masdar Institute Flagship Project, and the Tata Center for Technology and Design. The authors are with the Laboratory for Information and Decision Systems, Massachusetts Institute of Technology, Cambridge, MA. Emails: {dariam, mardavij, dahleh}@mit.edu

permanent battery service. Another difference is that the bounds in [6] are derived under a fixed initial energy allocation. Our results show that, at least for deferrable loads, this is restrictive. A battery model was also used in [12] to quantify the aggregate flexibility of a group of thermostatically controlled loads, and upper bounds on battery parameters were identified from data. In [11], an empirical method was developed to identify a lower bound on the battery capacity of a collection of flexible loads. The method is not restricted to a specific policy nor a particular load model.

Our work is also related to [13], where the authors characterize the *instantaneous* energy storage capacity of a collection of heterogeneous deferrable loads. This is different from our work, where we are interested in realizing *long-term* battery capacity. However, there is some overlap, which is explained in Remark 2.

*Notation*

For two vectors $x, y \in \mathbb{R}^3$, we say that $x \leq y$ if $x_i \leq y_i$ for all $i = 1, 2, 3$, and that $x < y$ if, in addition, $x_i < y_i$ for some $i$. We define the saturation operator as $[x]_a^b = \max(a, \min(x, b))$.

## II. DEFERRABLE ENERGY CONSUMPTION

A single deferrable energy load is characterized by an arrival time $\tau \in \mathbb{R}$, an energy demand, $E$, a time period, $T$, in which the demand must be filled, and a limit, $\overline{P}$, on its maximum power consumption. The energy consumed at time $t$, by a load with arrival time $\tau$ is denoted

$$e_\tau(t) = \int_{-\infty}^{t} p_\tau(\theta) d\theta, \quad (1)$$

where $p_\tau$ is the corresponding power consumption. The load admits any power consumption trajectory that satisfies its requirements, that is:

$$e_\tau(t) = 0, \text{ if } t \in (-\infty, \tau] \quad (2a)$$
$$e_\tau(t) = E \text{ if } t \in [\tau + T, \infty) \quad (2b)$$
$$0 \leq p_\tau(t) \leq \overline{P} \text{ for all } t, \tau \in \mathbb{R} \quad (2c)$$

We define the *nominal power consumption* of the load as

$$p_\tau^{\text{nom}}(t) = \begin{cases} P_0 & t \in [\tau, \tau + T] \\ 0 & \text{otherwise} \end{cases},$$

where $P_0 = E/T$. The flexibility of the load is due to its ability to deviate from nominal consumption. Figure 1 illustrates four possible energy trajectories for a flexible load.

Now consider a collection of deferrable loads that are indexed by their arrival times. We adopt the following assumptions.

- Arrival times are periodic with rate $\lambda > 0$, i.e., $\tau \in \mathbb{A}_\lambda = \{\ldots, -1/\lambda, 0, 1/\lambda, \ldots,\}$.
- All loads are identical, meaning that they have the same $E$, $T$, and $\overline{P}$.

We wish to use the aggregate flexibility of the loads to absorb exogenous power fluctuations, $w$, that are not known in advance. That is, we wish to select power trajectories, $p_\tau$, that satisfy (2) and

$$\sum_{\tau \in \mathbb{A}_\lambda} p_\tau(t) = \sum_{\tau \in \mathbb{A}_\lambda} p_\tau^{\text{nom}}(t) + w(t). \quad (3)$$

A scheduling policy, $\mu : w \to p$, decides how $w$ should be allocated over the loads. We let $p_\tau = \mu_\tau(w)$ denote the power allocated to load $\tau$. Since $w$ is uncertain, we restrict our attention to *causal* policies, that is, $\mu$ has no *a priori* knowledge of $w$. The set of uncertain power trajectories that can be absorbed by the loads under a given $\mu$ is denoted

$$\mathbb{W}_\lambda(\mu) = \{w : p_\tau = \mu_\tau(w) \text{ saisfies (2) and (3)}\}$$

To simplify the analysis, we will assume that the arrival rate $\lambda = \infty$. While a generalization of the results in Section IV to finite arrival rates can be obtained by using essentially the same techniques, it makes the analysis more cumbersome and does not add any significant understanding.

To account for infinite arrival rates, we introduce a modified notion of aggregate flexibility as follows. By $w \in \alpha \mathbb{W}_\lambda$, where $\alpha$ is a scalar, we mean that $w = \alpha w'$ for some $w' \in \mathbb{W}_\lambda$. We define

$$\mathbb{W}(\mu) = \lim_{\lambda \to \infty} \frac{1}{T\lambda} \mathbb{W}_\lambda(\mu)$$
$$= \left\{ w : \frac{1}{T} \int_{\mathbb{R}} \mu_\tau(w) d\tau = \frac{1}{T} \int_{\mathbb{R}} p_\tau^{\text{nom}} d\tau + w, \right.$$
$$\left. \mu_\tau(w) \text{ satisfies (2)} \right\} \quad (4)$$

For large $\lambda$, $T\lambda$ is an accurate estimate of the number of loads that have entered their active consumption phase. In this case, the set $\mathbb{W}(\mu)$ can be interpreted as the capacity of the average active load.

Let $x_\sigma(t) = e_{t-\sigma}(t)$ denote the energy of the load that arrived $\sigma$ seconds ago. We refer to $x_\sigma$, $\sigma \in [0, T]$ as the *energy allocation*. It is straightforward to show that

$$x_{\sigma+h}(t+h) = x_\sigma(t) + \int_{t}^{t+h} p_{t-\sigma}(\theta) d\theta \quad (5)$$

for $h \geq 0$. From (2) it follows that

$$\underline{x}_\sigma \leq x_\sigma(t) \leq \overline{x}_\sigma, \quad (6)$$

where $\overline{x}_\sigma = \left[\overline{P}\sigma\right]_0^E$ and $\underline{x}_\sigma = \left[\overline{P}(\sigma - (1 - \frac{P_0}{\overline{P}})T)\right]_0^E$. These bounds are illustrated by the gray lines in Figure 3 and are attained when all loads fill their demands as fast as possible, or defer their consumption for as long as possible.

The aggregate energy that is *stored* by the loads (in addition to their nominal energy) is denoted by

$$x_{\text{avg}}(t) = \frac{1}{T} \left( \int_0^T x_\sigma(t) d\sigma - \int_0^T x_\sigma^{\text{nom}} d\sigma \right), \quad (7)$$

where $x_\sigma^{\text{nom}} = [P_0 \sigma]_0^E$.

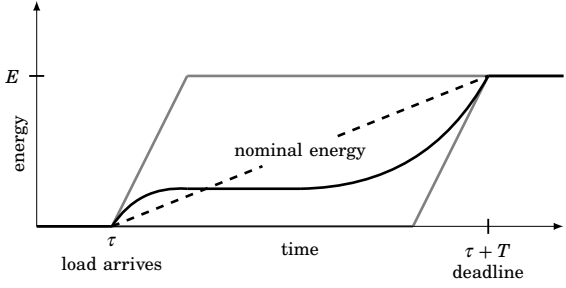

Fig. 1: Four energy trajectories for a deferrable load. All feasible trajectories must lie between the upper and lower gray curves, which illustrate the trajectories that fill the energy demand in the shortest time possible and postpone the consumption for as long as possible, respectively. The dashed line shows the nominal energy trajectory.

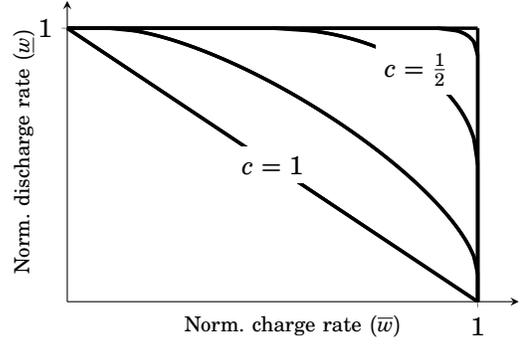

Fig. 2: Upper bounds on the normalized battery charge and and discharge rates that can be attained. The different curves correspond to normalized volume $c = \{0, 0.1, 0.5, 0.9, 1\}$.

## III. BATTERY EMULATION

An ideal battery is characterized by a volume, $C$, charge rate $\overline{W}$, and discharge rate $\underline{W}$. The exogenous power trajectories that can be absorbed by a battery with parameters $\phi = (C\ \overline{W}\ \underline{W})$ are given by

$$\mathbb{B}(\phi) = \left\{ w : -\underline{W} \leq w(t) \leq \overline{W},\ -\frac{C}{2} \leq \chi(t) \leq \frac{C}{2}, \right.$$
$$\left. \chi(t) = \int_{-\infty}^{t} w(\theta)d\theta,\ w(t) = 0 \text{ for } t < 0 \right\} \quad (8)$$

We say that a policy $\mu$ *realizes* the battery $\mathbb{B}(\phi)$ if $\mathbb{B}(\phi) \subset \mathbb{W}(\mu)$. The battery $\mathbb{B}(\phi)$ is called *realizable*, written $\mathbb{B}(\phi) \subset \mathbb{W}$, if there is *some* causal policy that realizes it.

Given load parameters $E$, $T$, $\overline{P}$, our goal is to characterize the set of batteries that can be realized.

*Remark 1 (Initialization phase):* As we show in Section IV, different initial energy allocations may be required for emulating different batteries. This means that, before starting the battery service, an initialization phase is required during which $\mu$ attains the required $x(0)$. For notational convenience, the initialization period is set to $(-\infty, 0)$.

## IV. BATTERY PARAMETER TRADE-OFFS

In this section, we show that there are fundamental trade-offs between the parameters of the batteries that can be emulated, and explain why this is the case. Proofs are presented in the appendix.

*Proposition 1:* Let $\phi_{\max} = (C_{\max}\ \overline{W}_{\max}\ \underline{W}_{\max})$, where

$$C_{\max} = E(1 - \tfrac{P_0}{\overline{P}}) \qquad \overline{W}_{\max} = \overline{P} - P_0 \qquad \underline{W}_{\max} = P_0.$$

Then $\mathbb{B}(\phi_{\max})$ is the smallest set that contains all $\mathbb{B}(\phi) \subset \mathbb{W}$. ▽

A consequence of Proposition 1 is that if there is no trade-off between battery parameters, that is, if there is a single largest realizable battery, it must be $\mathbb{B}(\phi_{\max})$. The next result shows that, in practice, this is not the case.

*Proposition 2:* $\mathbb{B}(\phi_{\max}) \subset \mathbb{W}$ if and only if $\overline{P} = \infty$. ▽

*Remark 2 (Overlap with previous work):* The bounds presented in [13, Section IV-B] imply that $\mathbb{B}(\phi_{\max})$ contains all realizable batteries, but do not establish tightness of this bound. Also, it follows from the results in [13, Section III] that if $\overline{P} = \infty$, then $\mathbb{B}(\phi_{\max})$ is realizable. This covers sufficiency in Proposition 2 but not necessity.

If we disregard the trivial case $\mathbb{B}(\phi) = \{0\}$ (for instance, this includes $\phi = (\infty\ 0\ 0)$) then $\phi \leq \phi_{\max}$ is a necessary condition for $\mathbb{B}(\phi) \subset \mathbb{W}$. Henceforth, we will assume:

$\mathcal{A}_1$: $\overline{P} < \infty$.

$\mathcal{A}_2$: $\phi \leq \phi_{\max}$

*Theorem 1:* Suppose $\mathcal{A}_{1\text{-}2}$ hold and set

$$c = C/C_{\max} \qquad \overline{w} = \overline{W}/\overline{W}_{\max} \qquad \underline{w} = \underline{W}/\underline{W}_{\max}.$$

Then $\mathbb{B}(\phi) \subset \mathbb{W}$ only if

$$(\overline{w} + \underline{w} - c)^2 \leq 4\overline{w}\underline{w}(1 - c), \quad (9)$$

whenever $\overline{w}, \underline{w} \geq 1 - c$ and $\overline{w} + \underline{w} \geq c$. ▽

Theorem 1 provides upper bounds on the capacity of realizable batteries. These bounds are depicted in Figure 2 for different energy volumes and show that any realizable $\mathbb{B}(\phi)$ must be substantially smaller that $\mathbb{B}(\phi_{\max})$. In particular, if we try to realize a battery where any two of the battery parameters are set to their individual bounds, the third must be zero. For instance, if we insist on a battery that is able to both charge at rate $\overline{W}_{\max}$ and discharge at rate $\underline{W}_{\max}$, then it will have zero volume.

*Remark 3 (Non-uniform rates):* A reason behind the substantial trade-offs in Theorem 1 is the requirement to guarantee *uniform* charge/discharge rates over all energy levels in $[-C/2, C/2]$. To better utilize the flexibility of the loads, one could replace $\mathbb{B}(\phi)$ with a different type of storage device where $\overline{W}$ and $\underline{W}$ depend on $\chi$. How to choose this dependence in order to best utilize the available flexibility is an interesting research direction.

The *charge slack*, $\overline{s}_\sigma$, and the *discharge slack*, $\underline{s}_\sigma$, are defined as

$$\overline{s}_\sigma = (E - x_\sigma)/\overline{P} \qquad (10)$$
$$\underline{s}_\sigma = T - \sigma - \overline{s}_\sigma \qquad (11)$$

respectively. For a load with energy level $x_\sigma$, the charge slack is the longest period it can maintain the maximum consumption rate, $\overline{P}$. It is a measure of the load's ability to *absorb* energy. Similarly, the discharge slack is the longest duration the load can maintain minimum, i.e., zero, consumption rate, and quantifies its ability to *release* energy. See Figure 3. Since $\overline{s}_\sigma + \underline{s}_\sigma = T - \sigma$, there is a conflict between maintaining a high $\overline{s}_\sigma$ and a high $\underline{s}_\sigma$.

The next results characterize the energy allocations under policies that emulate two illustrative batteries that satisfy (9).

*Proposition 3:* Let $\overline{\phi} = (C_{\max}, \overline{W}_{\max}, 0)$ and suppose that $\mathbb{B}(\overline{\phi}) \subset \mathbb{W}(\mu)$. Let $w \in \mathbb{B}(\overline{\phi})$ and set $p = \mu(w)$. Then the following implication holds

$$\overline{s}_\sigma(t) > \overline{s}_{\sigma'}(t) \implies x_\sigma(t) = \overline{x}_\sigma \text{ or } x_{\sigma'}(t) = \underline{x}_{\sigma'} \qquad (12)$$

for any $\sigma, \sigma' \in [0,T]$ and $t \in \mathbb{R}$. $\triangledown$

By Theorem 1, the battery $\mathbb{B}(\overline{\phi})$ is able to absorb the largest amount of energy faster than any other realizable battery. Proposition 3 states that the energy allocation under any policy that realizes $\mathbb{B}(\overline{\phi})$ must be as even as possible in terms of charge slacks. Similarly, the next result shows that any policy that realizes the battery that is best suited for releasing energy at a high rate must strive to equalize the discharge slacks.

*Proposition 4:* Let $\underline{\phi} = (C_{\max}, 0, \underline{W}_{\max})$ and suppose that $\mathbb{B}(\underline{\phi}) \subset \mathbb{W}(\mu)$. Let $w \in \mathbb{B}(\underline{\phi})$ and set $p = \mu(w)$. Then the following implication holds

$$\underline{s}_\sigma(t) > \underline{s}_{\sigma'}(t) \implies x_\sigma(t) = \underline{x}_\sigma \text{ or } x_{\sigma'}(t) = \overline{x}_{\sigma'} \qquad (13)$$

for any $\sigma, \sigma' \in [0,T]$ and $t \in \mathbb{R}$.. $\triangledown$

The energy allocations that satisfy (12) and (13) are shown in Figure 3. To understand why there is a trade-off between the batteries that can be emulated, suppose that $\mathbb{B}(\phi_{\max}) \subset \mathbb{W}(\mu)$ for some causal policy $\mu$, which, by Proposition 1, would be the case if there was no trade-off. Since $w$ is not known in advance, at time $t$, $\mu$ must be able to manage both of the following two outcomes: absorb energy at an aggregate rate $\overline{W}_{\max}$ until $x_{\text{avg}}(t') = C_{\max}/2$, and release energy at an aggregate rate $\underline{W}_{\max}$ until $x_{\text{avg}}(t'') = -C_{\max}/2$, for some $t'$, $t'' \geq t$. The first scenario requires *all* loads in their active consumption phase to consume at the maximum rate, $\overline{P}$, until *all* loads have reached their highest attainable energy level. This is only possible if $x(t)$ satisfies (12). The second scenario requires *all* loads in their active consumption phase to stop consumption until *all* loads have reached their lowest attainable energy level. This is only possible if $x(t)$ satisfies (13). Since it is not possible to instantaneously transition between energy allocations that satisfy (12) and (13), some battery capacity must be sacrificed. If we reduce $\overline{W}$ ($\underline{W}$) we relax the requirement that all of the active loads must be able to maintain their maximum (minimum) consumption rate for the duration it takes the battery to fully charge (discharge). If we reduce $C$, we relax the requirement that all loads must have reached their highest/lowest attainable energy levels for the battery to be considered fully charged/discharged.

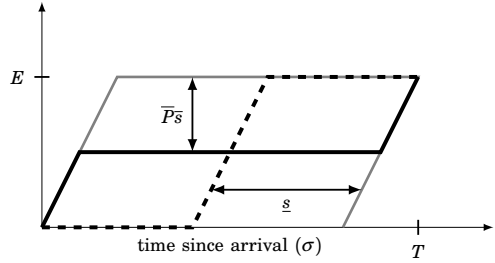

Fig. 3: Depiction of energy allocations, $x_\sigma(t)$, that satisfy (12) (solid) and (13) (dashed) for a fixed $t$. In both cases $x_{\text{avg}}(t) = 0$. The energy allocation (12) is better suited for tracking positive adjustments in the aggregate consumption rate, while (13) is better in terms of negative adjustments The gray curves show the maximum and minimum attainable energy levels $\overline{x}$ and $\underline{x}$.

The discussion above shows that there is a fundamental trade-off between the abilities of a collection of deferrable loads to absorb and release energy at high rates. This trade-off is not an artifact of our homogeneous and deterministic load model. It is a consequence of 1) the energy allocations required for absorbing and releasing energy at high rates have different slack profiles, 2) the dynamics of the loads prevent instantaneous transitions between these allocations, and 3) due to the uncertainty in the reference trajectory, $w$, it is not possible to infer the required energy allocation ahead of time. Since the definitions of charge and discharge slack naturally extend to non-identical loads, the identified trade-off occurs for collections of heterogeneous and uncertain deferrable loads as well.

## V. SUMMARY

We investigated the possibility of using the flexibility of a collection of deferrable loads to emulate a battery. A battery captures the three most important properties of energy storage: the volume of energy that can be stored, the rate at which it can be absorbed, and the rate at which it can be released.

Our main contribution was to derive upper bounds on the battery capacity that can be realized, and show that there is a fundamental trade-off between the abilities of the load collective to absorb and release energy at high aggregate rates. While these results were derived for a collection of homogeneous loads with periodic arrivals, our analysis shows that the identified trade-off occurs for collections of heterogeneous and uncertain loads as well.

## Appendix

We start by introducing some lemmas.

*Lemma 1:*
$$\frac{1}{T}\int_0^T (\overline{x}_\sigma - x_\sigma^{\text{nom}})\, d\sigma = \frac{1}{T}\int_0^T (x_\sigma^{\text{nom}} - \underline{x}_\sigma)\, d\sigma = \frac{C_{\max}}{2}$$

*Proof:* This follows immediately from the definitions of $\overline{x}$ and $\underline{x}$ (see Figure 3) and $x_\sigma^{\text{nom}} = P_0 \sigma$. ∎

If $w \in \mathbb{B}(\phi)$, the maximum duration the battery can maintain its maximum and minimum rates are
$$t_c(\chi) = \frac{C/2 - \chi}{\overline{W}} \quad \text{and} \quad t_d(\chi) = \frac{C/2 + \chi}{\underline{W}},$$
where the energy level, $\chi$, is defined in (8).

*Lemma 2:* Suppose that $\mathbb{B}(\phi) \subset \mathbb{W}(\mu)$ and let $p = \mu(w)$. Then, for each $w \in \mathbb{B}(\phi)$, there are functions $\underline{\delta}_\sigma, \overline{\delta}_\sigma : \mathbb{R} \to [0, \overline{E}]$, such that

$$\int_0^T \underline{\delta}_\sigma(t)d\sigma \leq \frac{T(C_{\max} - C)}{2} \tag{14}$$

$$\int_0^T \overline{\delta}_\sigma(t)d\sigma \leq \frac{T(C_{\max} - C)}{2} \tag{15}$$

$$\overline{z}_\sigma - \overline{\delta}_\sigma \leq x_\sigma \leq \underline{z}_\sigma + \underline{\delta}_\sigma \tag{16}$$

where

$$\overline{z}_\sigma(\chi) = \overline{x}_{\sigma+t_c(\chi)} - \overline{P}t_c(\chi) \quad \text{and} \quad \underline{z}(\chi) = \underline{x}_{\sigma+t_d(\chi)}. \tag{17}$$

*Proof:* Under the assumptions of the lemma, we have $x_{\text{avg}}(t) = \chi(t)$. Since $w$ is not known to $\mu$ a priori, at time $t$, $\mu$ must be able to handle both of the following outcomes
1) $w(\theta) = \overline{W}$ for $\theta \in [t, t+t_c]$
2) $w(\theta) = -\underline{W}$ for $\theta \in [t, t+t_d]$

These outcomes correspond to charging/discharging the battery at maximum rate until it is full/empty. If the outcome is 1), then $x_{\text{avg}}(t+t_c) = \chi(t+t_c) = C/2$. We can express $x(t+t_c) = \overline{x} - \delta$, for some $0 \leq \delta_\sigma \leq \overline{x}_\sigma - \underline{x}_\sigma \leq E$. Using this in (7), yields

$$\frac{C}{2} = \frac{1}{T}\int_0^T (x_\sigma(t+t_c) - x_\sigma^{\text{nom}})\, d\sigma$$
$$= \frac{1}{T}\int_0^T (\overline{x}_\sigma - \delta_\sigma - x_\sigma^{\text{nom}})\, d\sigma = \frac{C_{\max}}{2} - \frac{1}{T}\int_0^T \delta_\sigma d\sigma$$
$$\implies \frac{1}{T}\int_0^T \delta_\sigma d\sigma = \frac{(C_{\max} - C)}{2}$$

where the third equality follows from Lemma 1. From (5) and (2c) we get $\overline{x}_{\sigma+t_c} - \delta_{\sigma+t_c} = x_{\sigma+t_c}(t+t_c) \leq x_\sigma(t) + \overline{P}t_c$. Setting $\overline{\delta}_\sigma(t) = \delta_{\sigma+t_c}$, gives the first inequality in (16). Also, $\overline{\delta}_\sigma \in [0, E]$ and

$$\int_0^T \overline{\delta}_\sigma d\sigma = \int_{t_c}^{T+t_c} \delta_\sigma d\sigma = \int_{t_c}^T \delta_\sigma d\sigma \leq \frac{(C_{\max} - C)T}{2}$$

which establishes (15). The steps to verify (14) and the second inequality in (16) are analogous. ∎

### A. Proof of Proposition 1

Let $w \in \mathbb{W}(\mu)$ for some $\mu$ and set $p_\tau = \mu_\tau(w)$. Then

$$Tw(t) = \int_{t-T}^t p_\tau(t) - P_0 d\tau \leq (\overline{P} - P_0)T$$

and similarly we can show that $w(t) \geq -P_0$. The energy stored by the collection at time $t$ is $\int_0^t w(\theta)d\theta =$

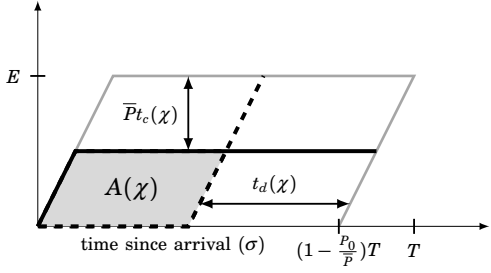

Fig. 4: The solid line shows $\bar{z}_\sigma$ for $\sigma \in [0, T - t_c]$ and the dashed line shows $\underline{z}_\sigma$ for $\sigma \in [0, T - t_d]$.

$x_{\text{avg}}(t)$, where $x_{\text{avg}}$ is defined by (7). Using that $x(t) \leq \bar{x}$ we have

$$\int_0^t w(\theta) d\theta \leq \frac{1}{T} \int_0^T (\bar{x}_\sigma - x_\sigma^{\text{nom}}) d\sigma = \frac{C_{\max}}{2},$$

where the last equality follows from Lemma 1. Similarly, we can show that $\int_{-\infty}^t w(\theta) d\theta \geq -C_{\max}/2$. Hence, $\mathbb{W}(\mu) \subset \mathbb{B}(\phi_{\max})$ and it follows that $\mathbb{B}(\phi_{\max})$ contains all realizable batteries.

To show that $\mathbb{B}(\phi_{\max})$ is the smallest set that contains all $\mathbb{B}(\phi) \subset \mathbb{W}$ it suffices to prove the implication

$$\mathbb{B}(\phi) \subset \mathbb{B}(\phi') \text{ for all } \mathbb{B}(\phi) \subset \mathbb{W} \implies \phi' \geq \phi_{\max}$$

It can be shown that the batteries in Proposition 3 and Proposition 4 are realizable (by policies that allocate as much power as possible to loads with the largest charge slack and smallest discharge slack, respectively). Hence, by assumption, $\mathbb{B}(\bar{\phi}) \subset \mathbb{B}(\phi')$ and $\mathbb{B}(\underline{\phi}) \subset \mathbb{B}(\phi')$. It follows that $\phi' \geq \bar{\phi}$ and $\phi' \geq \underline{\phi}$.

*B. Proof of Proposition 2*

Necessity follows from Theorem 1, which is established independently of Proposition 2. For the converse direction, we will show that if $\bar{P} = \infty$, then $\mathbb{B}(\phi_{\max}) \subset \mathbb{W}(\mu)$, where $\mu$ allocates power according to

$$x_\sigma(t) = \begin{cases} \underline{x}_\sigma & \text{if } \sigma \in [0, T - \varsigma(t)] \\ \bar{x}_\sigma & \text{if } \sigma \in (T - \varsigma(t), T] \end{cases} \quad (18)$$

and where $\varsigma(t) = \frac{C_{\max}/2 + \chi(t)}{W_{\max}}$. It is straightforward to verify that under (18) we have $x_{\text{avg}}(t) = \chi(t)$ for all $t$, so $p_\tau = \mu_\tau(w)$ satisfies

$$\frac{1}{T} \int_\mathbb{R} \mu_\tau(w) d\tau = \frac{1}{T} \int_\mathbb{R} p_\tau^{\text{nom}} d\tau + w \quad (19)$$

Also, as long as $\chi(t) \in [-\frac{C_{\max}}{2}, \frac{C_{\max}}{2}]$ we have $\varsigma(t) \in [0, T]$. Hence, $\mu_\tau(w)$ satisfies (2) for $w \in \mathbb{B}(\phi_{\max})$.

*C. Proof of Theorem 1*

By $\mathcal{A}_2$ it is sufficient to show that if $\mathbb{B}(\phi) \subset \mathbb{W}_\infty$ then (9) is satisfied on

$$\mathbb{S} = \{c, \bar{w}, \underline{w} \in [0, 1] : \bar{w}, \underline{w} \geq 1 - c, \bar{w} + \underline{w} \geq c\}.$$

The only point in $\mathbb{S}$ that corresponds to $\bar{w} = 0$ has $\underline{w} = 1$ and $c = 1$. This point satisfies (9). Similarly, condition (9) is also satisfied when $\underline{w} = 0$. Hence we may restrict our attention to $\bar{w}, \underline{w} > 0$.

It follows from Lemma 2 that for each $\chi \in [-C/2, C/2]$, there are non-negative scalars $\bar{\delta}_\sigma$ and $\underline{\delta}_\sigma$ that satisfy (15) and (14), such that

$$\bar{z}_\sigma(\chi) - \underline{z}_\sigma(\chi) \leq \bar{\delta}_\sigma + \underline{\delta}_\sigma. \quad (20)$$

Integrating both sides of this relation over $\sigma \in [0, T - t_c(\chi) - t_d(\chi)]$ yields

$$A(\chi) = \int_0^{T - t_c(\chi) - t_d(\chi)} \bar{z}_\sigma(\chi) - \underline{z}_\sigma(\chi) d\sigma \leq (C_{\max} - C)T \quad (21)$$

Since $C \leq C_{\max}$, the condition in (21) is satisfied for $\chi \in [-C/2, C/2]$ if and only if it is also satisfied on $\{\chi : A(\chi) \geq 0\}$. On this set, $A(\chi)$ equals the area of the parallelogram in Figure 4 with base and height

$$B(\chi) = (1 - \frac{P_0}{\bar{P}})T - t_d(\chi)$$
$$= \left(1 - \frac{P_0}{\bar{P}}\right) T \left(1 - \frac{C/2 + \chi}{\underline{w} C_{\max}}\right)$$
$$H(\chi) = E - \bar{P} t_c(\chi) = E \left(1 - \frac{C/2 - \chi}{\bar{w} C_{\max}}\right).$$

Hence, (21) holds on $\{\chi \in [-C/2, C/2] : A(\chi) > 0\}$ if and only if

$$a(\chi') = b(\chi') h(\chi') \leq 1 - c, \text{ for all } \chi' \in \mathbb{I}, \quad (22)$$

where $b(\chi) = (1 - \frac{c/2 + \chi'}{\underline{w}})$, $h(\chi') = (1 - \frac{c/2 - \chi'}{\bar{w}})$ and

$$\mathbb{I} = [-c/2, c/2] \cap \{\chi' : b(\chi'), h(\chi') \geq 0\}$$
$$= [-c/2, c/2] \cap [c/2 - \bar{w}, -c/2 + \underline{w}].$$

The extremum

$$\max_{\chi' \in \mathbb{R}} a(\chi') = \frac{(\bar{w} + \underline{w} - c)^2}{4 \bar{w} \underline{w}}$$

is attained at $\chi^* = (\underline{w} - \bar{w})/2$. The set $\mathbb{I}$ is non-empty if and only if $\bar{w} + \underline{w} \geq c$. In this case, it is straightforward to verify that $\chi^* \in \mathbb{I}$ if and only if $|\bar{w} - \underline{w}| \leq c$. We conclude that if $\mathbb{B}(\phi) \subset \mathbb{W}$ then (9) holds on

$$\{c, \bar{w}, \underline{w} \in [0, 1] : \bar{w} + \underline{w} \geq c, |\bar{w} - \underline{w}| \leq c\}.$$

This set contains $\mathbb{S}$.

*D. Proof of Proposition 3 (and Proposition 4)*

It follows from Lemma 2 that $x_\sigma \geq \bar{z}_\sigma(\chi)$. Combining this with (6) yields $x_\sigma \geq \max(\bar{z}_\sigma(\chi), \underline{x}_\sigma)$. Since, by assumption, $\chi = x_{\text{avg}}$, we have

$$\chi \geq \frac{1}{T} \int_0^{T-t_c} \bar{z}_\sigma - \underline{x}_\sigma d\sigma = -\frac{1}{T} \int_0^T x_\sigma^{\text{nom}} - \underline{x}_\sigma d\sigma$$
$$= (E - \bar{P} t_c)(1 - \frac{P_0}{\bar{P}}) - \frac{C_{\max}}{2} = \chi \quad (23)$$

so $x_\sigma = \max(\underline{z}_\sigma, \bar{x}_\sigma)$, which satisfies the assertion In the second to last equality in (23) we have used both Lemma 1 and $\int (\bar{z}_\sigma - \underline{x}_\sigma) = (E - \bar{P} t_c)(1 - \frac{P_0}{\bar{P}})T$, which is straightforward to derive using Figure 4. The proof of Proposition 4 is analogous.